\documentclass[aps,prd,twocolumn]{revtex4}

\usepackage{amssymb}
\usepackage{amsfonts}
\usepackage{amsmath}
\usepackage{graphicx}
\usepackage{bm}
\usepackage{color}
\usepackage[dvips]{epsfig}
\usepackage[dvips]{graphicx}
\usepackage{float}
\usepackage{epstopdf}
\usepackage[colorlinks=true,pdfstartview=FitV,linkcolor=blue,citecolor=blue,urlcolor=blue,breaklinks=true]{hyperref}

\begin{document}

\title{Self-dual configurations in Abelian Higgs models with $k$-generalized
gauge field dynamics}
\author{R. Casana$^1$}
\email{rodolfo.casana@gmail.com}
\author{A. Cavalcante$^1$}
\email{andre$_$cavs@hotmail.com}
\author{E. da Hora$^{1,2}$}
\email{edahora.ufma@gmail.com}
\affiliation{$^1${Departamento de F\'{\i}sica, Universidade Federal do Maranh\~{a}o,
65080-805, S\~{a}o Lu\'{\i}s, Maranh\~{a}o, Brazil.}\\
$^2$Coordenadoria Interdisciplinar de Ci\^{e}ncia e Tecnologia, Universidade
Federal do Maranh\~{a}o, {65080-805}, S\~{a}o Lu\'{\i}s, Maranh\~{a}o,
Brazil.}

\begin{abstract}
We have shown the existence of self-dual solutions in new Maxwell-Higgs scenarios
where the gauge field possesses a $k$-generalized dynamic, i.e., the
kinetic term of gauge field is a highly nonlinear function of $F_{\mu\nu}F^{\mu\nu}$.
We have implemented our proposal by means of a $k$-generalized model
displaying the spontaneous symmetry breaking phenomenon. We implement consistently the
Bogomol'nyi-Prasad-Sommerfield formalism providing highly nonlinear self-dual equations
whose solutions are electrically neutral possessing total energy proportional to the
magnetic flux. Among the infinite set of possible configurations, we have found families
of $k$-generalized models whose self-dual equations have a form mathematically similar
to the ones arising in the Maxwell-Higgs or Chern-Simons-Higgs models. Furthermore, we
have verified that our proposal also supports infinite twinlike models with  $|\phi
|^4$-potential or $|\phi |^6$-potential. With the aim to show explicitly that the BPS
equations are able to provide well-behaved configurations, we have considered a test
model in order to study axially symmetric vortices. By depending of the self-dual
potential, we have shown that the $k$-generalized model is able to produce solutions
that for long distances have a exponential decay (as Abrikosov-Nielsen-Olesen vortices)
or have a power-law decay (characterizing delocalized vortices). In all cases, we
observe that the generalization modifies the vortex core size, the magnetic field
amplitude and the bosonic masses but the total energy remains proportional to the
quantized magnetic flux.
\end{abstract}

\maketitle


\section{Introduction\label{Intro}}

 Configurations exhibiting nontrivial topology usually emerge as static
solutions of classical fields models presenting highly nonlinear
interactions. In general, this nonlinearity is introduced by means of a
potential describing the scalar-matter self-interaction \cite{n5}. Moreover,
the potential must allow the mechanism of the spontaneous symmetry breaking
happen, because it is known that the topological structures are formed
during symmetry-breaking phase transitions. It is the reason why these
solutions receive so much attention within the cosmological context or
condensed matter physics.

In the last years have received a special attention the topological objects
arising from noncanonical field models, some of them coming from string theories. These
new field theories (called $k$-generalized models) are characterized by possess
nonstandard kinetic terms that could play the role of a symmetry breaking potential
\cite{o1,MHgeral}. The new self-dual $k$-generalized solutions can exhibit similar
behavior as their conventional counterparts however the exotic dynamics can also
produce variations on the vortex-core size and on the field amplitudes \cite{o2}.
Furthermore, some $k$-models  possesses the same self-dual configurations (they have
exactly the same BPS equations), including their energy density and total energy than
their canonical counterparts, they are named are named \emph{twinlike models}
\cite{n007,n009,n008,n010}. Such a  versatility has motivated the use of nonstandard
kinetic terms in an attempt to explain the accelerated inflationary phase of the
universe \cite{n8}, strong gravitational waves \cite{sgw}, dark matter \cite{dm},
and others \cite{o}.

There are many scenarios motivating the study of the generalized
Maxwell-Higgs models. Among them we can cite the problem of the localization of
gravity and/or particle/field in a 6-D braneworld \cite{massimo} and the
Born-Infeld-Higgs models \cite{BI_1,Diego}. Another interesting scenario for to
study Maxwell-Higgs models is related to cosmic strings, topological defects
similar to the vortices, which may have great importance in the evolution of
the Cosmos and formation of the structures existing in the Universe
\cite{Vilenkin}. The community has renewed its interest because recently it was
established that superstrings theory supports cosmic strings solutions inside its
theoretical framework \cite{Sakellariadou}.

The aim of the present manuscript is go further in the study of self-dual
configurations in noncanonical models by considering a Maxwell-Higgs (MH) scenario
where the $k$-generalization is driven by the gauge  field kinetic
term,  i.e., the kinetic term of gauge field is a highly nonlinear function of
$F_{\mu\nu}F^{\mu\nu}$. We have organized our contribution as follows:
In Sec. \ref{general} we  establish the theoretical framework in which our studies
will be developed.  We implement consistently the Bogomol'nyi-Prasad-Sommerfield
formalism \cite{n4} providing highly nonlinear self-dual equations whose solutions are
electrically neutral possessing total energy proportional to the magnetic flux.
It was verified  $k$-generalized framework supports infinity twinlike self-dual models with $|\phi |^{4}$ or $|\phi|^{6}$ potentials. In Sec. \ref{mssb} we have managed to
establishing a class of models whose BPS equations become linear in the magnetic field,
i.e., they have mathematical expressions very similar to the self-dual equations
corresponding to the Maxwell-Higgs or Chern-Simons-Higgs models.   In Sec. \ref{axial},
we study the general properties of the axially symmetric vortex solutions generated
by an arbitrary $k$-generalized Abelian-Higgs model with $|\phi |^{4}$ and $|\phi|^{6}$
potentials. We have also propose a general potential generating self-dual delocalized vortex whose behavior for long distances is type a power decay. Finally, in Sec. \ref{end}, we present our ending comments and perspectives.

\section{Abelian Higgs models with $k$-generalized gauge field dynamics}

\label{general}

Our proposal consists of a possible generalized model where
the $k$-generalization idea is applied only to the dynamics of the gauge
field. We introduce an approach based in models supporting spontaneous
symmetry breaking potentials described by the following Lagrangian density,
\begin{equation}
\mathcal{L}=h(|\phi |)K(Y)+w(|\phi |)\left\vert D_{\mu }\phi \right\vert
^{2}-V(|\phi |),  \label{L_BIgeral}
\end{equation}%
where the complex scalar field $\phi$ stands for the Higgs one whose minimal
covariant derivative reading as $D_{\mu }\phi=\partial_{\mu}\phi-ieA_{\mu}\phi$,
with $A_{\mu}$ being the gauge field.  In the gauge field sector $h(|\phi |)$
is a nonnegative function and $K(Y)$ driving the $k$-generalization of the gauge
field is an arbitrary and nonpositive function of $Y$ which is defined by
\begin{equation}
Y\equiv -\frac{F_{\mu \nu }F^{\mu \nu }}{4\mathcal{U}(|\phi |)},  \label{yx}
\end{equation}%
being $F_{\mu \nu }=\partial_{\mu}A_{\nu}-\partial_{\nu}A_{\mu}$ the
electromagnetic field strength tensor. In the Higgs sector, $w(|\phi|) $
must be positive, whilst $V(|\phi |)$ represents some convenient potential.
The Lagrangian density (\ref{L_BIgeral}) allows to describe many
of the generalized models present in the literature, for example, we can
recover Born-Infeld-Higgs ones \cite{BI_1,Diego}, the prototype of highly
nonlinear gauge field dynamics.

In the remainder of the manuscript we are interested in the self-dual
configurations can be generated by these $k$-generalized models by
considering $w(|\phi |)=1$. Thus, we consider the $k$-generalized models
defined by the Lagrangian density,
\begin{equation}
\mathcal{L}=h(|\phi |)K(Y)+ \left\vert D_{\mu }\phi \right\vert ^{2}-V(|\phi
|).  \label{lg1}
\end{equation}
The equation of motion of the gauge field reads%
\begin{equation}
\partial _{\nu }\left( \frac{h}{\mathcal{U}}K_{Y}F^{\nu \mu }\right)
=eJ^{\mu },  \label{gg1}
\end{equation}%
where $K_{Y}\equiv dK/dY$ and\ $J_{\mu }$ is the conserved current density
given by%
\begin{equation}
J_{\mu }=i\left( \phi \partial _{\mu }\phi ^{\ast }-\phi ^{\ast }\partial
_{\mu }\phi \right) -2eA_{\mu }\left\vert \phi \right\vert ^{2}.  \label{crr}
\end{equation}

The equation of motion for the Higgs field is
\begin{equation}
D_{\mu }D^{\mu }\phi -K\frac{\partial h}{\partial \phi ^{\ast }}+\frac{h}{%
\mathcal{U}}YK_{Y}\frac{\partial \mathcal{U}}{\partial \phi ^{\ast }}+\frac{%
\partial V}{\partial \phi ^{\ast }}=0.  \label{gg2}
\end{equation}

From equation (\ref{gg1}) we obtain the Gauss and Amp\`ere laws for time
independent solutions,%
\begin{equation}
\partial _{j}\left( \frac{h}{\mathcal{U}}K_{Y}\partial _{j}A_{0}\right)
=2e^{2}A_{0}\left\vert \phi \right\vert ^{2}, \\
\end{equation}%
\begin{equation}
\epsilon _{kj}\partial _{j}\left( \frac{h}{\mathcal{U}}K_{Y}B\right) =eJ_{k},
\label{A_glx0}
\end{equation}%
respectively.

From the Gauss law, the electric charge density is given by $J_0=-2e{2}A_{0}\left\vert \phi \right\vert ^{2}$ whose integral provides  null total electric charge,
\begin{equation}
Q=\int d^2x J_0=0.
\end{equation}
It can be proved directly by integrating the Gauss law under suitable boundary conditions, i.e., for $|x|\rightarrow \infty$, $A_0\rightarrow 0$ and $|\phi|\rightarrow cte.$ Therefore, the field configurations are electrically neutral.

We observe that for such configurations the gauge condition $A_{0}=0$ is compatible
with the Gauss. Consequently, for this gauge condition, the Higgs field equation reduces to be%
\begin{equation}
D_{k}D_{k}\phi +K\frac{\partial h}{\partial \phi ^{\ast }}-\frac{h}{\mathcal{%
U}}YK_{Y}\frac{\partial \mathcal{U}}{\partial \phi ^{\ast }}-\frac{\partial V%
}{\partial \phi ^{\ast }}=0.  \label{H_glx1}
\end{equation}

\subsection{The BPS formalism}

The energy-momentum tensor for the models described by the Lagrangian
density (\ref{lg1}) is
\begin{eqnarray}
T_{\mu \nu } &=&-\frac{h}{\mathcal{U}}K_{Y}F_{\mu \beta }F_{\nu }{}^{\beta
}+(D_{\mu }\phi )^{\ast }D_{\nu }\phi \\[0.15cm]
&&+(D_{\nu }\phi )^{\ast }D_{\mu }\phi -\eta _{\mu \nu }\mathcal{L}.  \notag
\end{eqnarray}%
The energy is given by the integration of the {$T_{00}$} component which for
time-independent configurations with $A_{0}=0$ reads,
\begin{equation}
E=\int d^{2}x\left[ \frac{{}}{{}}-hK+\left\vert D_{k}\phi \right\vert ^{2}+V%
\right] .  \label{etotal}
\end{equation}%
It is positive-definite due to the conditions previously imposed to the functions
$h(|\phi |)$ and $K(Y)$.

We begin the implementation of the BPS formalism \cite{n4} by introducing the identity
\begin{equation}
\left\vert D_{k}\phi \right\vert ^{2}=\left\vert D_{\pm }\phi \right\vert
^{2}\pm eB\left\vert \phi \right\vert ^{2}\pm \frac{1}{2}\epsilon
_{ik}\partial _{i}J_{k},  \label{ssdd}
\end{equation}%
into the total energy (\ref{etotal}) that after some algebraic
manipulations can expressed as
\begin{eqnarray}
E &=&\int d^{2}x\left[ \left\vert D_{\pm }\phi \right\vert ^{2}+\frac{%
\mathcal{U}}{2hK_{Y}}\left( \frac{h}{\mathcal{U}}K_{Y}B\mp \sqrt{W}\right)
^{2}\right.  \notag \\[0.15cm]
&&\hspace{1.25cm}\pm B\left( \sqrt{W}+e\left\vert \phi \right\vert
^{2}\right) \pm \frac{1}{2}\epsilon _{ik}\partial _{i}J_{k}  \label{enn} \\%
[0.15cm]
&&\hspace{1.25cm}\left. -\frac{\mathcal{U}}{2hK_{Y}}W-\frac{1}{2}\frac{h}{%
\mathcal{U}}K_{Y}B^{2}-hK+V\right] ,  \notag
\end{eqnarray}
where we have introduced the function $W(|\phi |)$ that will be determined later.
At this point, we remember that the  BPS formalism consist in expressing the energy
density as a sum of quadratic terms plus a term proportional to the magnetic field
plus a total derivative. In our case, such requirements are achieved in two steps: The
first one consists in to determine $W(|\phi |)$ by choosing that the factor
multiplying the magnetic field in (\ref{enn}) to be $ev^2=cte.$, thus, we
get
\begin{equation}
W=e^{2}\left( v^{2}-\left\vert \phi \right\vert ^{2}\right) ^{2}.  \label{w1}
\end{equation}%
A second step is to require that the third row in (\ref{enn}) be null,
\begin{equation}
\frac{\mathcal{U}}{2hK_{Y}}W+\frac{1}{2}\frac{h}{\mathcal{U}}%
K_{Y}B^{2}+hK-V=0,  \label{xx10}
\end{equation}%
establishing a relation between all the functions defining the generalized
model. Importantly that the Eq. (\ref{xx10}) it is not arbitrary because, as
we will see later, in the BPS limit, it becomes equivalent to the condition:
$T_{11}+T_{22}=0$, proposed by Schaposnik and Vega \cite{SV} to obtain
self-dual configurations.

By fulfilling the conditions (\ref{w1}) and (\ref{xx10}), the energy (\ref%
{enn}) is written as
\begin{eqnarray}
E &=&\int d^{2}x\left\{ \pm ev^{2}B\pm \frac{1}{2}\epsilon _{ik}\partial
_{i}J_{k}+\left\vert D_{\pm }\phi \right\vert ^{2}\frac{{}}{{}}\right.
\label{xxq0} \\[0.15cm]
&&\hspace{1.1cm}\left. +\frac{\mathcal{U}}{2hK_{Y}}\left[ \frac{h}{\mathcal{U%
}}K_{Y}B\mp e\left( v^{2}-\left\vert \phi \right\vert ^{2}\right) \right]
^{2}\right\} .  \notag
\end{eqnarray}

The integration of the first term in Eq. (\ref{xxq0}) provides the total
magnetic flux,%
\begin{equation}
\int d^{2}xB=\Phi ,
\end{equation}%
and, under suitable boundary conditions, the integration of the total
derivative in Eq. (\ref{xxq0}) gives null contribution to the energy.

Then, from Eq. (\ref{xxq0}), we can see that total energy has a lower bound
\begin{equation}
E\geq \pm ev^{2}\Phi ,  \label{lower1}
\end{equation}%
which is attained by field configurations satisfying the BPS or self-dual
equations,%
\begin{equation}
D_{\pm }\phi =0,  \label{bps_n}
\end{equation}%
\begin{equation}
\frac{h}{\mathcal{U}}K_{Y}B=\pm e\left( v^{2}-\left\vert \phi \right\vert
^{2}\right) .  \label{bps_n0}
\end{equation}%
Besides, the Eqs. (\ref{xx10}) in the BPS limit becomes
\begin{equation}
\frac{h}{\mathcal{U}}K_{Y}B^{2}+hK-V=0,  \label{sv0}
\end{equation}%
which corresponds exactly to the condition $T_{11}+T_{22}=0$, from the
Schaposnik and Vega formalism. It also allows us to determine explicitly the
self-dual potential,
\begin{equation}
V_{sd}=\frac{h}{\mathcal{U}}K_{Y}B^{2}+hK\geq 0,  \label{sv1}
\end{equation}%
where $K_{Y}$ must be a positive function.

Finally, from Eq. (\ref{etotal}) we obtain the energy density of the solutions
satisfying the \emph{Bogomol'nyi bound} (\ref{lower1}), i. e., the field
configurations are solutions the self-dual equations (\ref{bps_n}) and (\ref{bps_n0}),
\begin{equation}
\varepsilon _{_{BPS}}=\frac{h}{\mathcal{U}}K_{Y}B^{2}+|D_{k}\phi |^{2}.
\label{e_bps}
\end{equation}
It will be positive-definite since the functions $h(|\phi|) $, $\mathcal{U}(|\phi |)$
and $K_{Y}(Y)$ also are.

It can be verified that the BPS equations (\ref{bps_n}) and (\ref{bps_n0}) also solve
the second-order Euler-Lagrange equations given by Eqs. (\ref{A_glx0}) and
(\ref{H_glx1}) when the potential (\ref{sv1}) is considered.

\subsection{Self-dual configurations with $k$-generalized gauge field
dynamics without an explicit SSB potential}

The results obtained for the model (\ref{lg1}) can be used to
analyze the interesting case when there is no an explicit spontaneous
symmetry breaking potential. Such a situation is obtained by setting $%
V(|\phi |)=0$ and $h(|\phi |)=\mathcal{U}(|\phi |)$ resulting in the
following Lagrangian density,
\begin{equation}
\mathcal{L}=\mathcal{U}(|\phi |)K(Y)+\left\vert D_{\mu }\phi \right\vert
^{2}.  \label{1x}
\end{equation}%
Obviously, the equations of motion for both gauge and Higgs
fields are modified, but the configurations remain electrically neutral.
The self-dual configurations have the total energy (\ref{lower1}) and they
are described by the following BPS or self-dual equations: The first one  is exactly the
Eq. (\ref{bps_n}) and the second one is obtained from (\ref{bps_n0}) to be
\begin{equation}
K_{Y}B =\pm e\left( v^{2}-\left\vert \phi \right\vert ^{2}\right) .
\label{bps_n2}
\end{equation}%
Similarly to the Eq.(\ref{bps_n0}), it can be highly nonlinear
in the magnetic field $B(|\phi |)$ due to the presence of $K_{Y}(Y)$.

Besides, the Eq. (\ref{sv0}) is reduced to the following form
\begin{equation}
K_{Y}B^{2}+K\mathcal{U}_{sd}=0,  \label{svx1}
\end{equation}
where $\mathcal{U}_{sd}$ means the self-dual form of the
function $\mathcal{U}(|\phi |)$, which can be explicitly determined
\begin{equation}
\mathcal{U}_{sd}(|\phi |)=-\frac{2U_{_{MH}}(|\phi |)}{K(Y)K_{Y}(Y)}.
\label{BPSpot2}
\end{equation}%
The function $U_{_{MH}}(|\phi |)$ express the usual self-dual
potential of the Maxwell-Higgs model given by
\begin{equation}
U_{_{MH}}(|\phi |)=\frac{e^{2}}{2}\left( v^{2}-\left\vert \phi \right\vert
^{2}\right) ^{2}.  \label{upx}
\end{equation}

From Eq. (\ref{BPSpot2}) we can see that function $\mathcal{U}%
_{sd}(|\phi|)$ would the role of a self-dual potential in the $k$%
-generalized model (\ref{1x}) with no an explicit spontaneous symmetry
breaking potential.

\subsection{Twinlike self-dual models}

In this section, our purpose is to show that the Lagrangian density
(\ref{lg1}) supports different models with the same self-dual configurations (they have
exactly the same BPS equations), including their energy density and total energy,
they are called twinlike models.

\subsubsection{$|\protect\phi|^4$-twinlike self-dual models}

Twinlike models to Maxwell-Higgs model must have the second
BPS equation written as follows
\begin{equation}
B=\pm e\left( v^{2}-\left\vert \phi \right\vert ^{2}\right) ,  \label{bpsn2}
\end{equation}%
So, the models described by the Lagrangian density (\ref{lg1})
attain this condition if we impose the following constraint in the BPS
equation (\ref{bps_n0}),
\begin{equation}
hK_{Y}=\mathcal{U}.  \label{twin1}
\end{equation}%
Consequently, from Eq. (\ref{sv1}) we determine the self-dual
potential to be
\begin{equation}
V_{sd}=2U_{_{MH}}+hK,
\end{equation}
and from Eq. (\ref{e_bps}) the BPS energy density reads
\begin{equation}
\varepsilon _{_{BPS}}=2U_{_{MH}}+|D_{k}\phi |^{2},
\end{equation}%
that is exactly the one of the Maxwell-Higgs model with
its corresponding total energy given by the lower bound (\ref{lower1}).

We observe that the BPS equations (\ref{bps_n}) and (\ref
{bpsn2}) do not depend explicitly of the functions $h( |\phi|) $, $K_Y(Y)$
and $\mathcal{U}(|\phi|)$. This allows to conclude that there are an
infinite number of twinlike models to Maxwell-Higgs model, one for each
chosen set of the generalizing functions whenever the condition (\ref{twin1}
) is satisfied.

\subsubsection{$|\protect\phi|^6$-twinlike self-dual models}

We will show that the Lagrangian density (\ref{lg1}) also
supports twinlike models possessing self-dual solutions electrically neutral
whose BPS equations are exactly the same as the ones of the Chern-Simons-Higgs
model. It is possible to be achieved if the generalizing functions in Eq.
(\ref{bps_n0}) satisfy the following condition
\begin{equation}
hK_{Y}=\frac{\kappa ^{2}}{2e^{2}}\frac{\mathcal{U}}{|\phi |^{2}}.
\label{cst}
\end{equation}%
Thus, the BPS equation (\ref{bps_n0}) is written in the exact
form like that of the Chern-Simons-Higgs model \cite{Hong,Lee},
\begin{equation}
B=\pm \frac{2e^{3}}{\kappa ^{2}}\left\vert \phi \right\vert ^{2}\left(
v^{2}-\left\vert \phi \right\vert ^{2}\right).
\end{equation}%
Then, from (\ref{sv1}) we obtain the self-dual potential
\begin{equation}
V_{sd}=2U_{_{CSH}}+hK\geq 0,
\end{equation}%
where $U_{_{CSH}}$ reads for the self-dual potential of the
Chern-Simons-Higgs model
\begin{equation}
U_{_{CSH}}(\left\vert \phi \right\vert )=\frac{e^{4}}{\kappa ^{2}}\left\vert
\phi \right\vert ^{2}(v^{2}-\left\vert \phi \right\vert ^{2})^{2}.
\label{pmcs}
\end{equation}%
Finally, from the Eq. (\ref{e_bps}), the BPS energy density becomes
\begin{equation}
\varepsilon _{_{BPS}}=2U_{_{CSH}}+|D_{k}\phi |^{2},
\end{equation}
that is exactly the one of the CSH model.

Similarly to previous case, we conclude that there are infinite
models possessing configurations with exactly the same BPS equations, BPS energy density and total energy than the Chern-Simons-Higgs electrodynamics but, in this case, the
self-dual configurations are electrically neutral. Again, there is a model
for each chosen set of generalizing functions satisfying the condition (\ref{cst}).

\section{Some $k$-generalized models \label{mssb}}

We will show the existence of $k$-generalized models described by the
Lagrangian density (\ref{lg1}) whose  BPS equations (\ref{bps_n}) and (\ref{bps_n0})
becomes  linear in the magnetic field such as it happens in the case of the
Maxwell-Higgs or Chern-Simons-Higgs models,  i.e., $|D_{\pm}\phi|=0$
and $B=\pm \mathcal{V}(V_{sd}(|\phi|),h(|\phi|),{\cal{U}}(|\phi|)$, where $\mathcal{V}$
is some function of the self-dual potential and generalizing functions. Among these
models we can include the twinlike ones shown in previous section.

\subsection{A simplest model\label{m1_1}}

A model given the simplest solution of the BPS equations is defined by
\begin{equation}
K(Y)=-\alpha \left( -Y\right) ^{\beta },  \label{ss0}
\end{equation}%
with $\alpha,\beta>0$ being constant parameters.

The second BPS equation (\ref{bps_n0}) becomes,
\begin{equation}
B=\pm \frac{2\beta }{2\beta -1}\frac{V_{sd}}{\sqrt{2U_{_{MH}}}},  \label{ss1}
\end{equation}%
where we must consider $\beta >1/2$, the potential $U_{_{MH}}(|\phi|)$ is given
by Eq. (\ref{upx}) and $V_{sd}(|\phi|)$ stands by the self-dual potential computed
from Eq. (\ref{sv1}),
\begin{equation}
V_{sd}{(|\phi |)=}\left[ \frac{{(2\beta -1)}^{2\beta -1}}{{\alpha \beta
^{2\beta }}}\right] ^{\frac{1}{2\beta -1}}\left( \frac{\mathcal{U}U_{_{MH}}}{%
{h}^{1/\beta }}\right) ^{\frac{\beta }{2\beta -1}},  \label{pot_x1}
\end{equation}
it is expressed in terms of the arbitrary functions $h(|\phi |)$ and
$\mathcal{U}(|\phi |)$.

Therefore, the BPS can be fixed whether we know the functions $%
h(|\phi |)$ and $\mathcal{U}(|\phi |)$ or we give the explicit form of the
self-dual potential. In the case when we fix the self-dual potential and $%
h(|\phi |)$ or $\mathcal{U}(|\phi |)$ it is possible to determine the other.

For example, by choosing $V_{sd}(|\phi |)=U_{_{MH}}(|\phi |)$,
a $|\phi|^4$-potential, the BPS equation (\ref{ss1}) reads
\begin{equation}
B=\pm \frac{\beta }{2\beta -1}e\left( v^{2}-\left\vert \phi \right\vert
^{2}\right) ,  \label{4m22}
\end{equation}%
which is similar to the Maxwell-Higgs one.

On the other hand, by selecting the $|\phi |^{6}$-potential given in (\ref{pmcs}),
$V_{sd}{(|\phi |)} =U_{_{CSH}}(|\phi |)$, the equation (\ref{ss1}) becomes
\begin{equation}
B=\pm \frac{\beta }{2\beta -1}\frac{2e^{3}}{\kappa ^{2}}\left\vert \phi
\right\vert ^{2}\left( v^{2}-\left\vert \phi \right\vert ^{2}\right) ,
\label{6m22}
\end{equation}%
which looks like with the one of the Chern-Simons-Higgs model \cite{Hong,Lee}.

\subsection{Other simplest models\label{m1_2}}

Other class of models providing simple self-dual equations can
be obtained by imposing in the BPS limit the following condition
\begin{equation}
\frac{B^{2}}{2\mathcal{U}}=\Delta =cte,  \label{cdx1}
\end{equation}%
with $\Delta >0$. It fixes $Y=-\Delta $ and, consequently, both $K(\Delta)<0
$ and $K_{Y}(\Delta )>0$ become well-defined constants.

In the following, we will show that the constant $\Delta $ and
the function $\mathcal{U}(|\phi |)$ determine completely the BPS structure
of the model. First, by using (\ref{cdx1}) in Eq. (\ref{bps_n0}) we obtain
the function $h(|\phi |)$ as
\begin{equation}
h(|\phi |)=\frac{\sqrt{\Delta \mathcal{U}U_{_{MH}}}}{\Delta K_{Y}(\Delta )}.
\label{hhh}
\end{equation}
Next, the Eq. (\ref{cdx1}) together with the Eq. (\ref{hhh})
simplify enormously the Eq. (\ref{sv1}) such that the self-dual potential
becomes
\begin{equation}
V_{sd}(|\phi |)=\frac{2\Delta K_{Y}(\Delta )+K(\Delta )}{\Delta K_{Y}(\Delta
)}\sqrt{\Delta \mathcal{U}U_{_{MH}}},  \label{VSD}
\end{equation}%
it will be nonnegative whenever we choose a function $K(Y)$
satisfying $2\Delta K_{Y}(\Delta )+K(\Delta )>0$. Finally, the BPS equation (%
\ref{bps_n0}) can be expressed in terms of the self-dual potential
\begin{equation}
B=\pm \frac{2\Delta K_{Y}(\Delta )}{2\Delta K_{Y}(\Delta )+K(\Delta )}\frac{%
V_{sd}}{\sqrt{2U_{_{MH}}}}.  \label{bpsm2}
\end{equation}

We conclude that the condition (\ref{cdx1}) beside to determine
completely the BPS structure of the model also generates a infinity family
of self-dual configurations described by fixing $\mathcal{U}(|\phi |)$ and
running $\Delta $. Alternatively, we can observe that the Eq. (\ref{VSD})
also allows to fix the function $\mathcal{U}(|\phi |)$ if we know the
self-dual potential $V_{sd}(|\phi|)$.

For example, we consider the self-dual models described by the $|\phi |^{4}$%
-potential (\ref{upx}), $V_{sd}(|\phi |)=U_{_{MH}}(|\phi |)$, the BPS
equation (\ref{bpsm2}) reads
\begin{equation}
B=\pm \frac{\Delta K_{Y}(\Delta )}{2\Delta K_{Y}(\Delta )+K(\Delta )}e\left(
v^{2}-\left\vert \phi \right\vert ^{2}\right) ,  \label{4m1-2}
\end{equation}%
it turns on very similar to the one of the Maxwell-Higgs model.

We can also consider the self-dual models described by the $\left\vert \phi
\right\vert ^{6}$-potential (\ref{pmcs}), i.e., $V_{sd}(|\phi |)=U_{_{CSH}}
(|\phi |)$, in this case BPS equation (\ref{bpsm2}) is
\begin{equation}
B=\pm \frac{\Delta K_{Y}(\Delta )}{2\Delta K_{Y}(\Delta )+K(\Delta )}
\frac{2e^{3}}{\kappa ^{2}}\left\vert \phi \right\vert ^{2}(v^{2}-
\left\vert \phi \right\vert ^{2}),  \label{6m1-2}
\end{equation}%
which is analogue to the one of the Chern-Simons-Higgs model but in the present
case the self-dual solutions are electrically neutral.

\subsection{$k$-generalized models without explicit SSB potential \label{nossb}}

Such as it happens in the general case described by Lagrangian density (\ref{lg1})
it is possible to show that the Lagrangian density (\ref{1x}) also supports the
existence of $k$-generalized models whose second BPS equation (\ref{bps_n2}) is
linear in the magnetic field. Such models can be obtained by imposing at the BPS
limit the following condition:
\begin{equation}
\frac{B^{2}}{2\mathcal{U}_{sd}}=\Delta >0,  \label{ccddt}
\end{equation}%
it is similar to the Eq. (\ref{cdx1}) but now the function $\mathcal{U}_{sd}(|\phi |)$
stands for the self-dual form obtained from Eq. (\ref{svx1}). By using the above
condition in Eq. (\ref{svx1}) allows to write
\begin{equation}
K(\Delta )=-2\Delta K_{Y}(\Delta ).
\end{equation}
This condition allows to find the all possible values of $\Delta$ for a
given model defined by the function $K(Y)$.

Thus, the second BPS equation (\ref{bps_n2}) and the self-dual
self-interaction $\mathcal{U}_{sd}(|\phi |)$ become
\begin{eqnarray}
B&=&\pm \frac{e}{K_{Y}(\Delta )}\left( v^{2}-\left\vert \phi \right\vert
^{2}\right) ,  \label{m2-2} \\[0.15cm]
\mathcal{U}_{sd}(|\phi |)&=&\frac{1}{\Delta \left[ K_{Y}(\Delta )\right] ^{2}%
}\frac{e^2}{2}\left(v^2-|\phi|^2\right)^2,  \label{sd_pt}
\end{eqnarray}
which now mathematically look very similar to the ones of the Maxwell-Higgs
model.

It is importantly to note that the condition (\ref{ccddt}) only provides  self-dual
models supporting $|\phi |^{4}$-self-interaction. It is a remarkable difference with
the simplest models described in the previous cases despite the conditions
(\ref{cdx1}) and (\ref{ccddt}) are very similar.

For example, a model satisfying all above conditions is given by
\begin{equation}
K(Y )=-\alpha(-Y)^{1/2},
\end{equation}
with $\alpha>0$.

\section{The self-dual vortex solutions \label{axial}}

In this section, we seek axially symmetric solutions according to the usual
vortex \emph{Ansatz} \cite{NO}
\begin{eqnarray}
\phi (r,\theta ) &=& vg(r)e^{in\theta }, \\[0.15cm]
A_{\theta }(r,\theta )&=&-\frac{a(r)-n}{er},  \label{ax2}
\end{eqnarray}
with $n=\pm 1,\pm 2,\pm 3...$ standing for the winding number of the vortex
solutions.  The profiles $g(r)$ and $a(r)$ are regular functions
describing solutions possessing finite energy and obeying the boundary
conditions,
\begin{eqnarray}
g(0) &=&0,\;\;a(0)=n,  \label{bcx1} \\[0.2cm]
g(\infty ) &=&1\text{\textbf{,}}\;\;a(\infty )=0.  \label{bcx2}
\end{eqnarray}
The magnetic field is given by
\begin{equation}
B(r)=-\frac{1}{er}\frac{da}{dr},  \label{B}
\end{equation}
and the self-dual energy density (\ref{e_bps}) reads as
\begin{equation}
\varepsilon _{sd}=\frac{h}{\mathcal{U}}K_{Y}B^{2}+2v^{2}\left( \frac{ag}{r}
\right) ^{2}.  \label{enn2}
\end{equation}
The total energy of the self-dual solutions is given by the lower bound (\ref{lower1}),
\begin{equation}
E_{sd}=\pm ev^{2}\Phi _{B}=\pm 2\pi v^{2}n,  \label{bx}
\end{equation}
it is proportional to the winding number of the vortex solution, as expected.

\subsection{Vortices in ${|\protect\phi|^4}$-models\label{phi4-mod}}

As we have seen all self-dual models have the same first BPS equation (\ref{bps_n}).
The difference is in the second BPS equation (\ref{bps_n0}) which depends of the
specific model to be analyzed. For the $|\phi |^{4}$-models here analyzed, the
BPS (\ref{bps_n0}) can be given by Eq. (\ref{4m22}) or Eq. (\ref{4m1-2}) or Eq.
(\ref{m2-2}). All of them can be written in a unique form:
\begin{equation}
\frac{dg}{dr}=\pm \frac{ag}{r},  \label{bpsx02}
\end{equation}%
\begin{equation}
B=-\frac{1}{er}\frac{da}{dr}=\pm f_{_{4}}ev^{2}(1-g^{2}),  \label{bpsx2}
\end{equation}
where the parameter $f_{_{4}}>0$ (depending on $\beta$ or $\Delta $) is
\begin{equation}
f_{_{4}}=\left\{
\begin{array}{ccc}
\displaystyle{\frac{\beta }{2\beta -1}} & , & \text{in Eq. (\ref{4m22}), } \\[0.4cm]
\displaystyle\frac{\Delta K_{Y}(\Delta )}{2\Delta K_{Y}(\Delta )+K(\Delta )}
& , & \text{in Eq. (\ref{4m1-2}),} \\[0.35cm]
\displaystyle\frac{1}{K_{Y}(\Delta )} & , & \text{in Eq. (\ref{m2-2}).}
\end{array}
\right.
\end{equation}%

In addition, the self-dual energy density is given by
\begin{equation}
\varepsilon _{sd}=2f_{_{4}}{U_{_{MH}}}+2v^{2}\left( \frac{ag}{r}\right) ^{2}.
\label{denxt}
\end{equation}

\begin{figure}[H]
\centering\includegraphics[width=8.6cm]{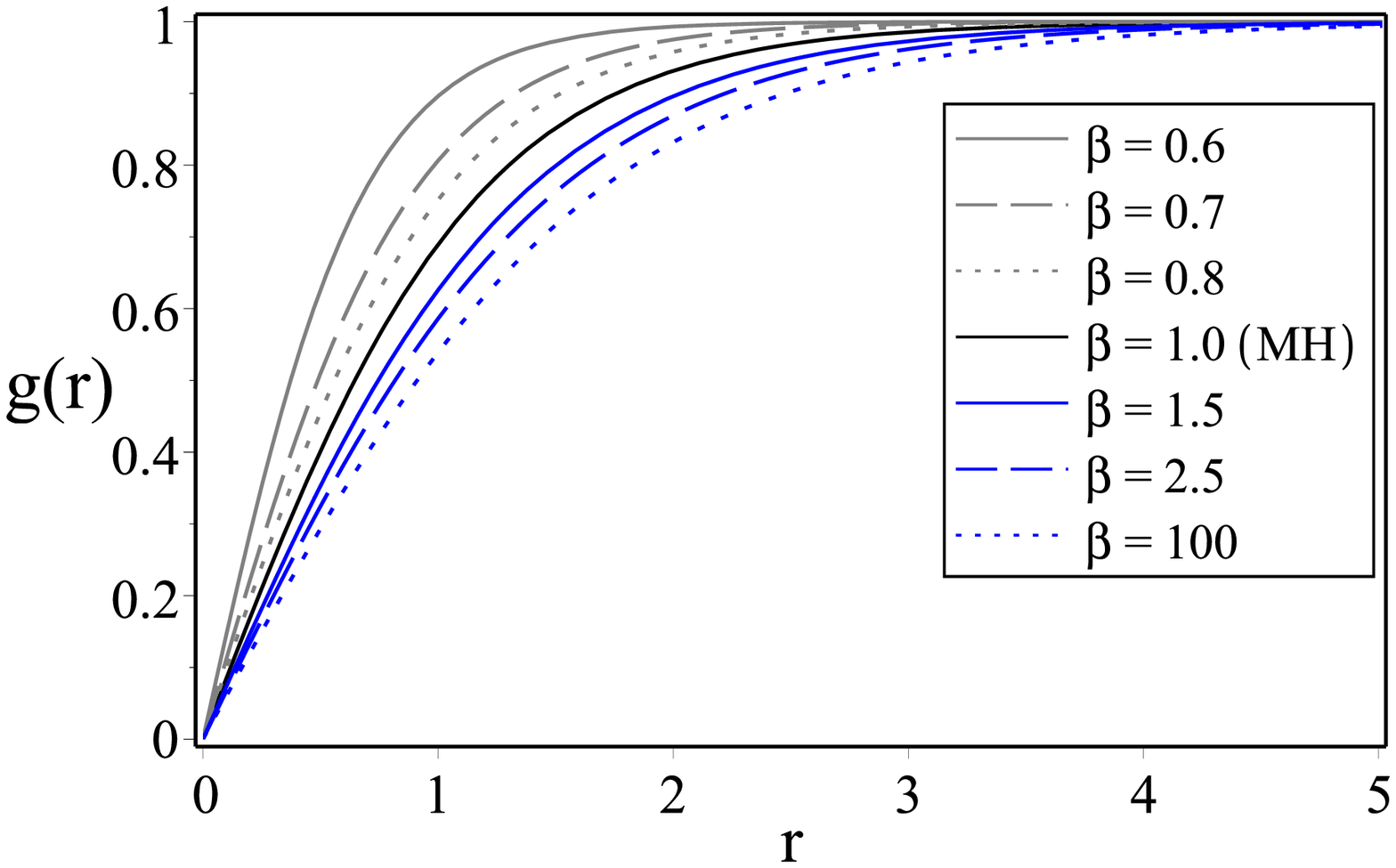}
\centering\includegraphics[width=8.6cm]{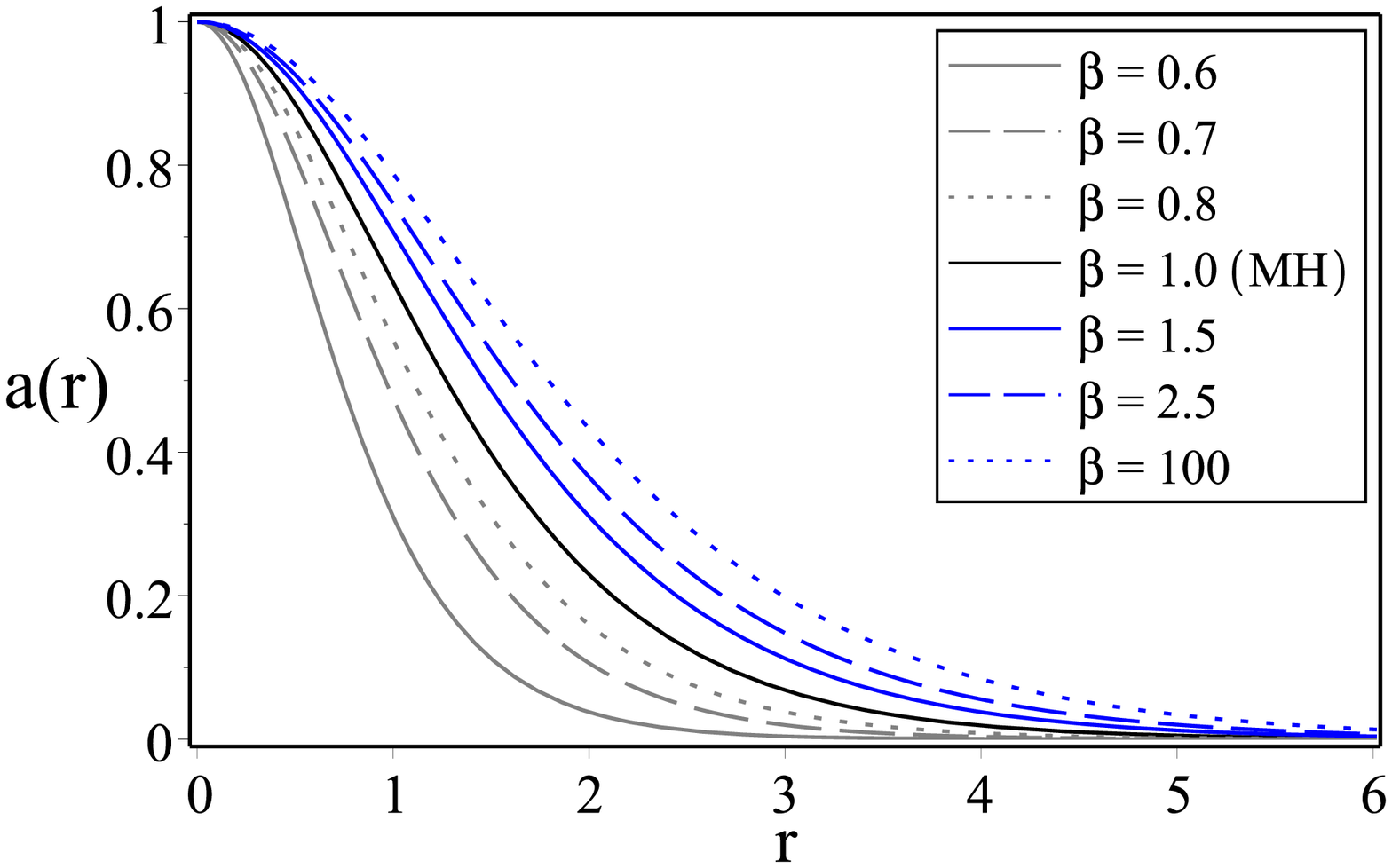}
\caption{The Higgs profile $g(r)$ (upper) and the gauge field profile $a(r) $ (lower)
for the model (\protect\ref{ss0}) with a $|\protect\phi|^4$-self-dual-potential (%
\protect\ref{upx}) and some values of $\protect\beta>1/2$.}
\label{fig01}
\end{figure}

The behavior of $g(r)$ and $a(r)$ near the boundaries can be easily determined by
solving the self-dual equations (\ref{bpsx02}) and (\ref{bpsx2}) around the
boundary values (\ref{bcx1}) and (\ref{bcx2}). This way, near the origin,
the profile functions behave as
\begin{eqnarray}
g(r) &\approx &G_{n}^{^{(f_{_{4}})}}r^{\left\vert n\right\vert }+..., \\%
[0.15cm]
a(r) &\approx &n\mp \frac{e^{2}v^{2}}{2}f_{_{4}}r^{2}+....
\end{eqnarray}%
On the other hand, when $r\rightarrow \infty $ they behave as
\begin{eqnarray}
g(r) &\simeq &1-\frac{G_{\infty }^{^{(f_{_{4}})}}}{\sqrt{r}}e^{-m_{_{4}}r},
\\[0.15cm]
a(r) &\simeq &{m_{_{4}}}G_{\infty }^{^{(f_{_{4}})}}\sqrt{r}e^{-m_{_{4}}r}.
\end{eqnarray}%
The constants $G_{n}^{^{(f_{_{4}})}}>0$ and $G_{\infty }^{^{(f_{_{4}})}}$
can be determined only numerically and $m_{_{4}}$ being the self-dual mass,
\begin{equation}
m_{_{4}}=m_{_{MH}}\sqrt{f_{_{4}}},  \label{M_G}
\end{equation}%
with $m_{_{MH}}=\sqrt{2}ev$ standing for the mass of the usual self-dual
Maxwell-Higgs bosons.

\begin{figure}[H]
\centering\includegraphics[width=8.6cm]{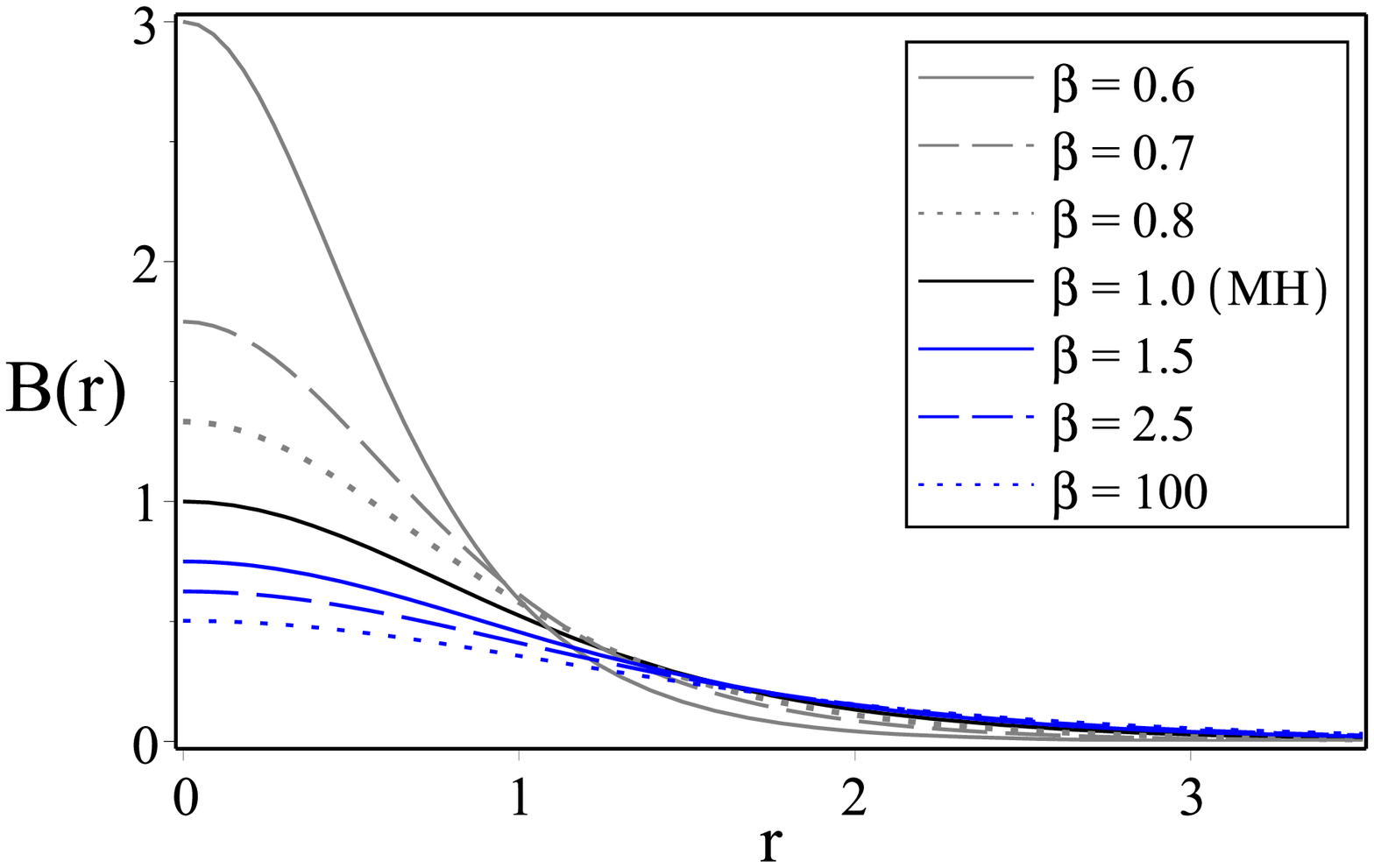}
\centering\includegraphics[width=8.6cm]{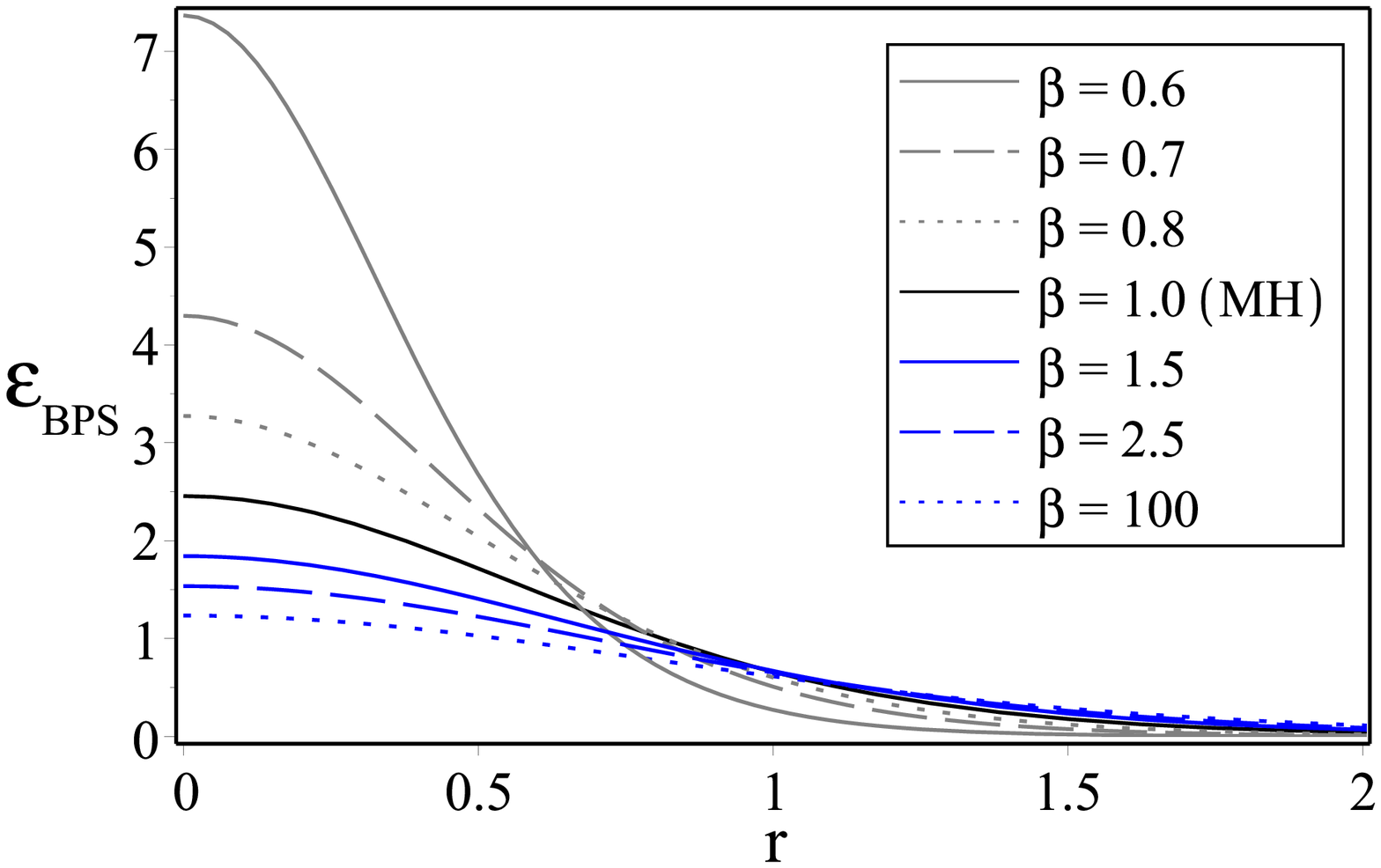}
\caption{The magnetic field $B(r) $ (upper) and the BPS energy density $
\protect\varepsilon_{_{BPS}}(r) $ (lower) coming from the model (\protect\ref
{ss0}) with a $|\protect\phi|^4$-self-dual-potential (\protect\ref{upx}) and
some values of $\protect\beta>1/2$.}
\label{fig02}
\end{figure}

We perform the numerical analysis of the $|\phi|^4$-models arising from the
$k$-generalized model defined in Eq. (\ref{ss0}), i.e., we consider the BPS
equations (\ref{bpsx02}) and (\ref{bpsx2}) with
\begin{equation}
 f^{(\beta)}_{_{4}}=\frac{\beta}{2\beta-1}, \label{ff4}
\end{equation}
for some values of $\beta> 1/2$.

In order to compute the numerical solutions we choose
the upper signs, $e = v=1$ and the configurations with winding number $n = 1$. The
profiles for the Higgs and gauge fields are given in Fig. \ref{fig01}, the correspondent
ones for the  magnetic field and the self-dual energy density are depicted in  Fig.
\ref{fig02}.

A brief analysis of the amplitude (\ref{ff4}) elucidates that, within the range
$1/2<\beta<1 $, the mass $m_{_{4}} $ increase for $\beta\rightarrow 1/2$, whilst
reaching $m_{_{4}}\rightarrow m_{_{MH}}$ for $\beta\rightarrow 1$. On the other hand,
for $\beta>1$, the mass $m_{_{4}}$ decreases continuously whenever $\beta$ increases attaining its minimum value $m_{_{4}}\rightarrow ev$ when   $\beta\rightarrow\infty$.
It explains the changes in the vortex-core size showed in the Figs. \ref{fig01} and \ref{fig02}.

For $n=1$, the magnetic field and the BPS energy density attain their maximum
amplitude at origin (see Fig. \ref{fig02}) and they are given by
\begin{eqnarray}
B(0)&=& ev^2 f^{(\beta)}_{_{4}}, \label{B0}\\[0.15cm]
\varepsilon _{_{BPS}}(0) &=& e^{2}v^{4}f^{(\beta)}_{_{4}}+2v^{2}(G_{1}^{^{(f_{_{4}})}})^{2},
\end{eqnarray}
respectively. The function (\ref{ff4}) explains clearly because the amplitudes
in relation to the MH ones ($\beta=1$) are higher for $1/2<\beta<1$ or smaller
for $\beta>1$.

From the Eqs. (\ref{bpsx2}), (\ref{M_G}) and (\ref{denxt}) we can establish the
general conclusion for $|\phi|^4$-configurations when compared with the
correspondent Maxwell-Higgs vortices: The $k$-generalization besides modifies
the amplitude of magnetic field and the masses of the self-dual bosons also alters
the self-dual energy density. The change in the boson mass value implies that the
vortex-core size can be increased or diminished.

\subsection{Vortices in ${|\protect\phi |^{6}}$-models}

We have demonstrated that the Lagrangian density (\ref{lg1}) also supports
$|\phi |^{6}$-models whose self-dual equations are given by (\ref{bps_n}) together
with (\ref{6m22}) or (\ref{6m1-2}).  Such equations in the vortex \emph{Ansatz} are written as
\begin{equation}
\frac{dg}{dr}=\pm \frac{ag}{r},  \label{6bpsx1}
\end{equation}
\begin{equation}
B=-\frac{1}{er}\frac{da}{dr}=\pm f_{_{6}}\frac{2e^{3}v^{4}}{\kappa ^{2}}%
g^{2}(1-g^{2}),  \label{6bpsx2}
\end{equation}%
where  $f_{_{6}}>0$ (depending on $\beta $ or $\Delta $)$\ $is
given by
\begin{equation}
f_{_{6}}=\left\{
\begin{array}{ccc}
\displaystyle{\frac{\beta }{2\beta -1}} & , & \text{in Eq. (\ref{6m22}), } \\
[0.35cm]
\displaystyle\frac{\Delta K_{Y}(\Delta )}{2\Delta K_{Y}(\Delta )+K(\Delta )}
& , & \text{in Eq. (\ref{6m1-2}).}
\end{array}
\right.
\end{equation}

The self-dual energy density is given by
\begin{equation}
\varepsilon _{sd}={2f_{_{6}}U_{_{CSH}}}+2v^{2}\left( \frac{ag}{r}\right)
^{2}.  \label{denxt1}
\end{equation}

The profiles for $r\rightarrow 0$ behave as
\begin{eqnarray}
g(r) &\approx &G_{n}^{^{(f_{_{6}})}}r^{\left\vert n\right\vert }+..., \\%
[0.15cm]
a(r) &\approx &n\mp f_{_{6}}\frac{e^{4}v^{4}}{\kappa ^{2}}\frac{\left(
G_{n}^{^{(f_{_{6}})}}\right) ^{2}}{\left( |n|+1\right) }r^{2|n|+2}+....
\end{eqnarray}%
For $r\rightarrow \infty $, the asymptotic behavior is
\begin{eqnarray}
g(r) &\simeq &1-\frac{G_{\infty }^{^{(f_{_{6}})}}}{\sqrt{r}}e^{-m_{_{6}}r},\\[0.15cm]
a(r) &\simeq &\frac{G_{\infty }^{^{(f_{_{6}})}}}{m_{_{6}}}\sqrt{r}e^{-m_{6}r}.
\end{eqnarray}%
The constants $G_{n}^{^{(f_{_{6}})}}>0$ and $G_{\infty }^{^{(f_{_{6}})}}$
are determined  numerically. The self-dual mass $m_{_{6}}$ is given by
\begin{equation}
m_{_{6}}=m_{_{CSH}}\sqrt{f_{_{6}}},  \label{mass}
\end{equation}%
with $m_{_{CSH}}=2e^{2}v^{2}/\kappa $ standing for self-dual CSH mass.

\begin{figure}[H]
\centering\includegraphics[width=8.6cm]{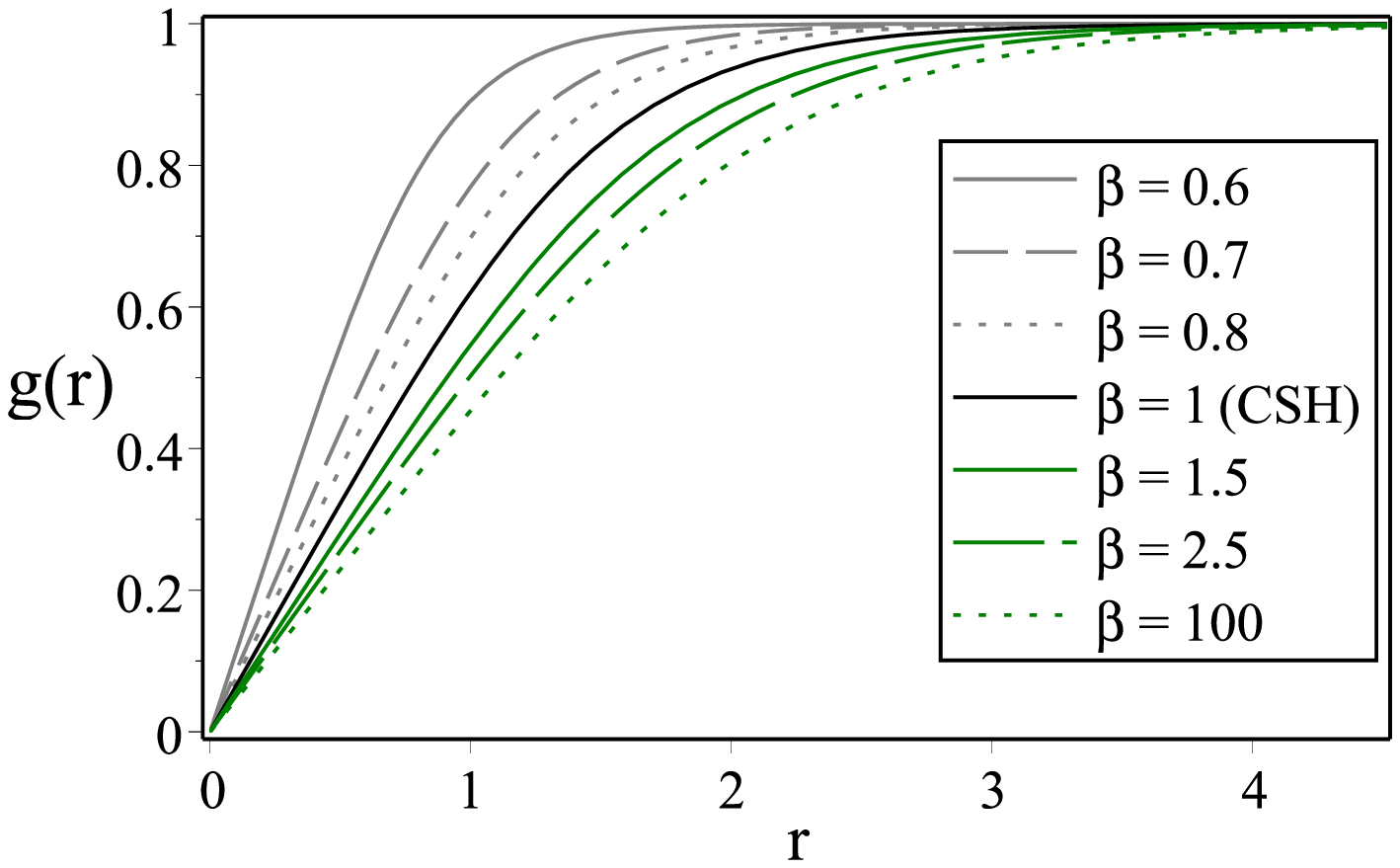} \centering\includegraphics[width=8.6cm]{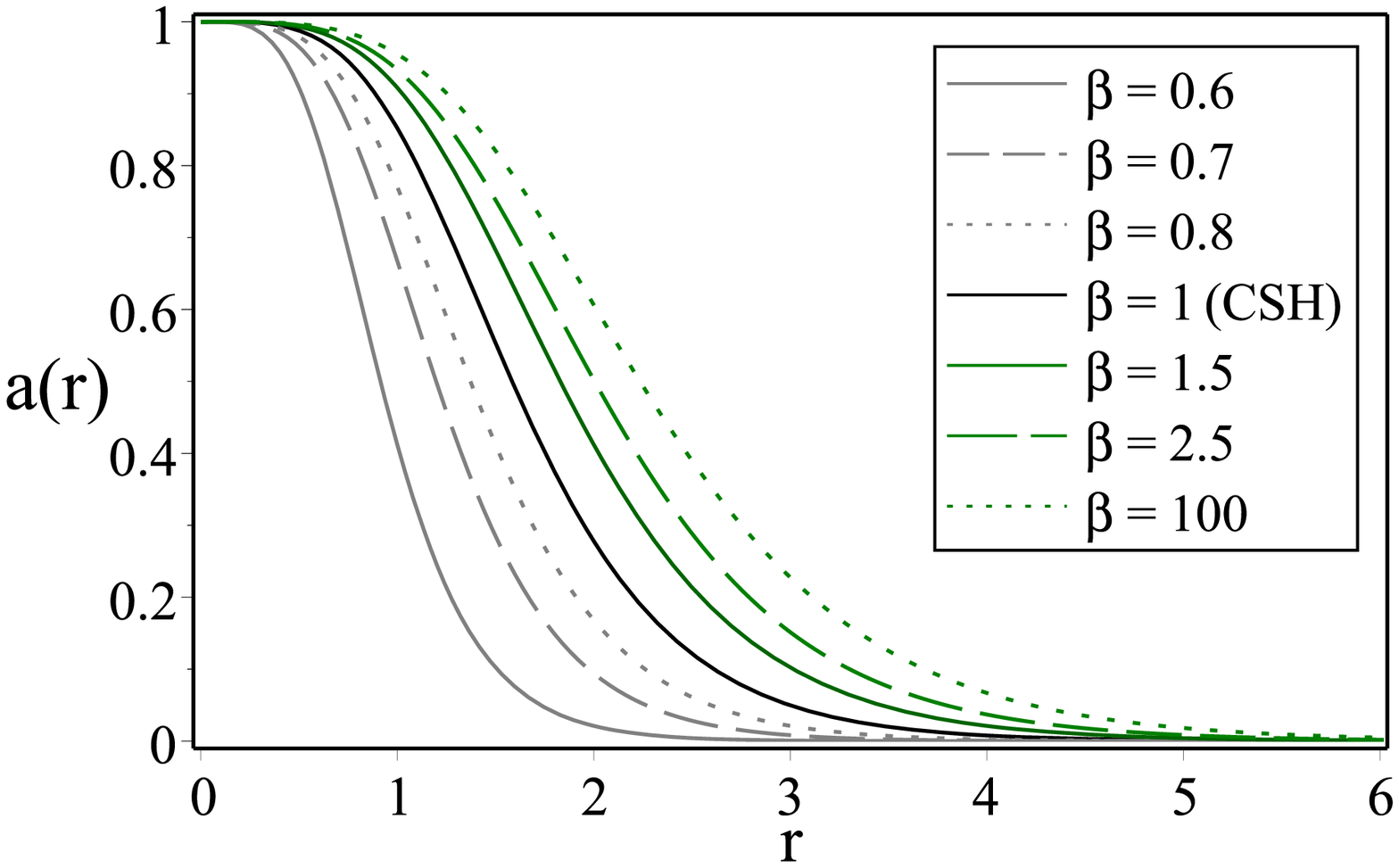}
\caption{The profiles $g(r) $ (upper) and $a(r) $ (lower) coming from the
model (\protect\ref{ss0}) with the $|\protect\phi|^6$-self-dual-potential (\protect\ref{pmcs}) and some values of $\protect\beta>1/2$.}
\label{fig03}
\end{figure}

\begin{figure}[H]
\centering\includegraphics[width=8.6cm]{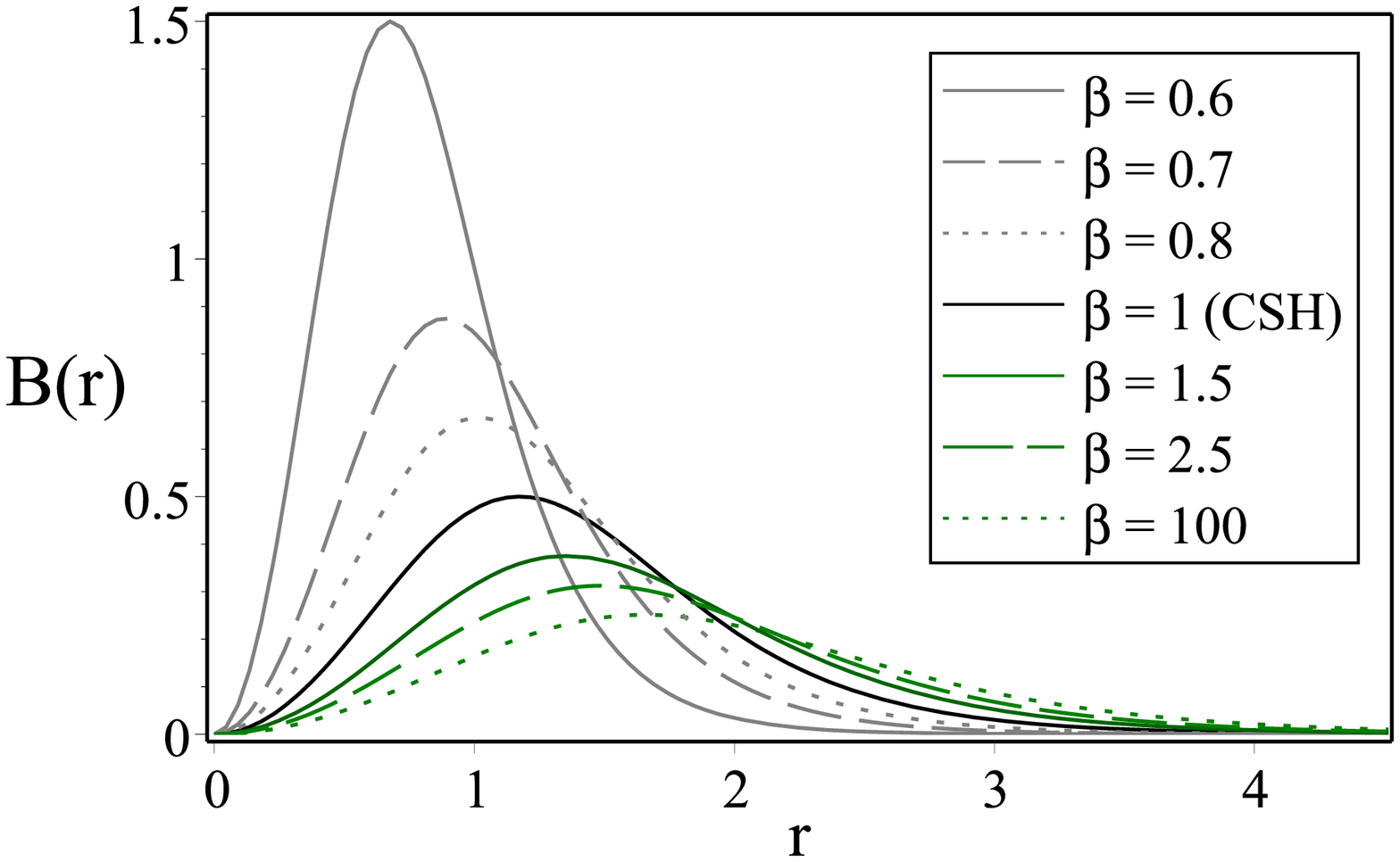}
\centering\includegraphics[width=8.6cm]{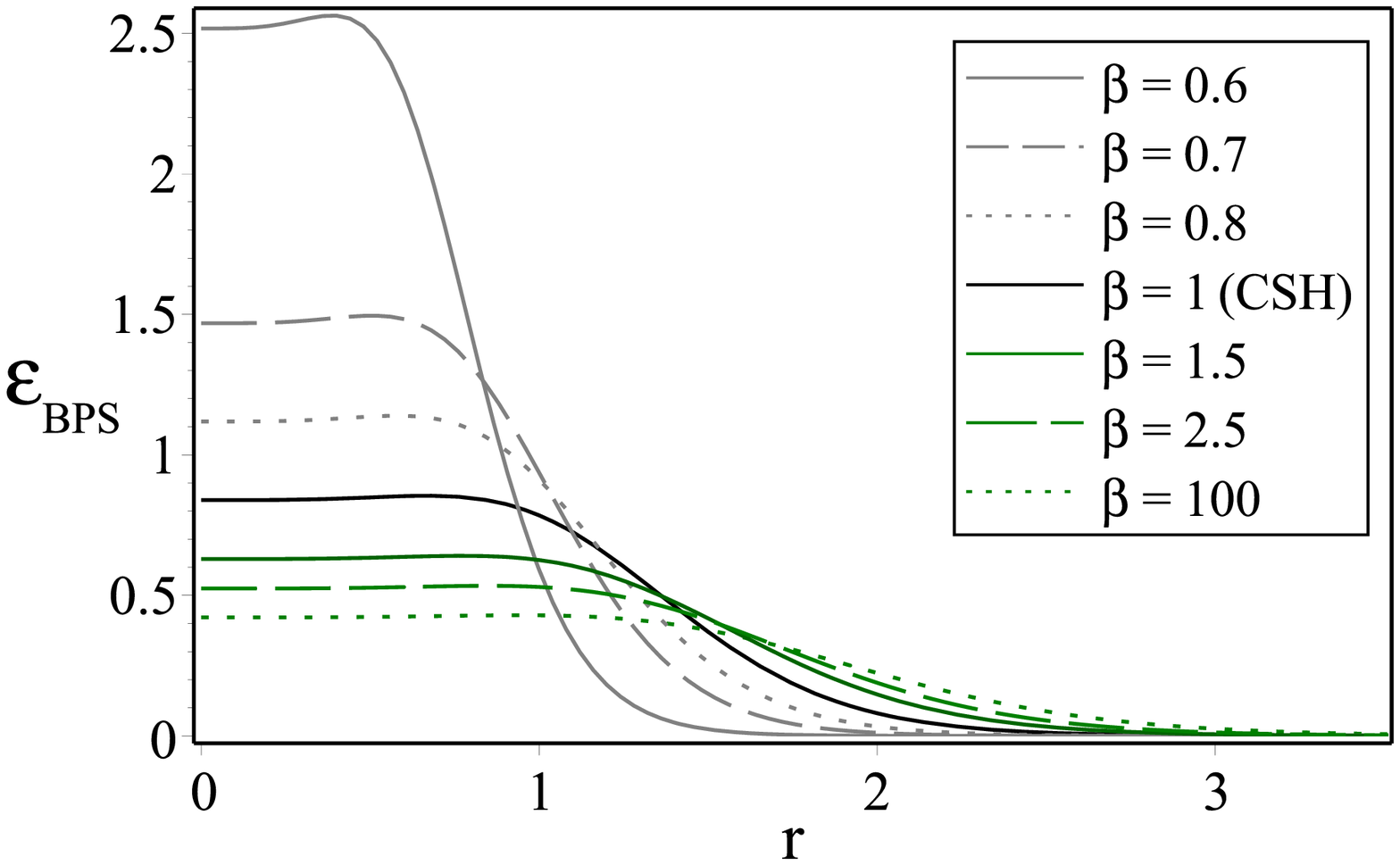}
\caption{The magnetic field $B(r) $ (upper) and the BPS energy density $%
\protect\varepsilon_{_{BPS}}(r) $ (lower) coming from the model (\protect\ref%
{ss0}) with the $|\protect\phi|^6$-self-dual-potential (\protect\ref{pmcs}) and
some values of $\protect\beta>1/2$.}
\label{fig04}
\end{figure}

Similarly to the case of the $|\phi|^4$-models, we perform the numerical
analysis of the $|\phi|^6$-models arising from the $k$-generalized model defined in Eq.
(\ref{ss0}). So, we consider the BPS equations (\ref{bpsx02}) and (\ref{bpsx2}) with
$f^{ (\beta)}_{_{6}}$ given by Eq. (\ref{ff4}), i.e.,
$f^{(\beta)}_{_{6}}=\beta/(2\beta-1)$ . In order to compute the numerical
solutions we choose the upper signs, $e = v=\kappa=1$ and the configurations with
winding number $n=1$. The profiles for the Higgs and gauge fields are given in Fig.
\ref{fig03}, the correspondent ones for the  magnetic field and the self-dual energy
density are depicted in  Fig. \ref{fig04}.

The Figs. \ref{fig03} and \ref{fig04} shows the changes in the core-size of the
vortex whenever the parameter $\beta$ changes its values. Such a effect has been
also observed in the $|\phi|^4$-models analyzed previously.

For $n=1$, the magnetic field profiles are rings  around the origin (see upper figure
in Fig. \ref{fig04}) whose maximum amplitude is
\begin{eqnarray}
B^{^{(\text{max})}}(r^*)=\frac{e^{3}v^{4}}{2\kappa^{2}}\frac{ \beta}{2\beta-1} ,
\end{eqnarray}
for a $r^*$ such that $g(r^*)=\sqrt{2}/2$.

On the other hand, for $n=1$, the amplitude at origin of the BPS energy density is
\begin{eqnarray}
\varepsilon _{_{BPS}}(0) = 2v^{2}(G_{1}^{^{(f_{_{6}})}})^{2},
\end{eqnarray}
it changes whenever $\beta$ do  it, i.e., the amplitudes in relation to $\beta=1$
ones are higher for $1/2<\beta<1$ or smaller for $\beta>1$ (see lower figure in Fig. \ref{fig04}).

\subsection{Delocalized self-dual vortices}

In the previous examples we have studied localized vortex solutions whose behavior
for very large values of $r$ is similar to the one of the Abrikosov-Nielsen-Olesen
vortices, i.e., they have a exponential decay. Now we go to show that the
$k$-generalized models defined by the Lagrangian density (\ref{lg1}) can engender
delocalized vortex, i.e., solutions possessing  for $r\rightarrow\infty$ a power-law
decay.

We present such solution by means of the model defined in Eq. (\ref{ss0})
whose second BPS equation is given by Eq. (\ref{ss1}). The delocalized vortices are
obtained by choosing the following self-dual potential,
\begin{equation}
V_{sd}(g)=\frac{e^{2}v^{4}}{2}\left( 1-g^{2}\right) ^{2+2/\gamma}.  \label{Vsd}
\end{equation}
It is obtained from (\ref{pot_x1}) by selecting appropriately the functions $h(g)$
and  $\mathcal{U}(g)$.  Here the parameter $\gamma>0$ will define the
power-law decay ($r^{-\gamma}$) for large values of $r$ of the self-dual solutions.

\begin{figure}[H]
\centering\includegraphics[width=8.6cm]{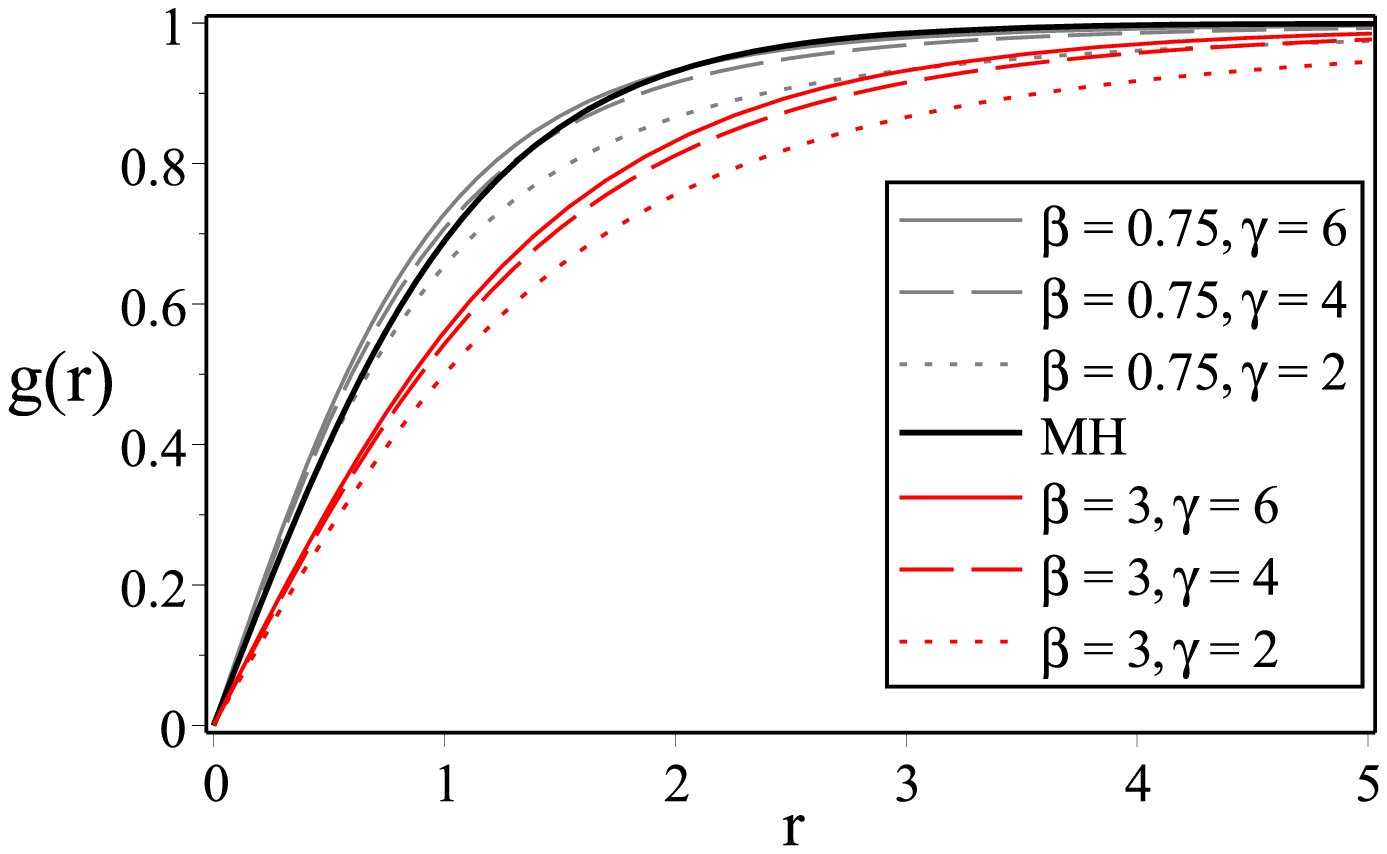}
\centering\includegraphics[width=8.6cm]{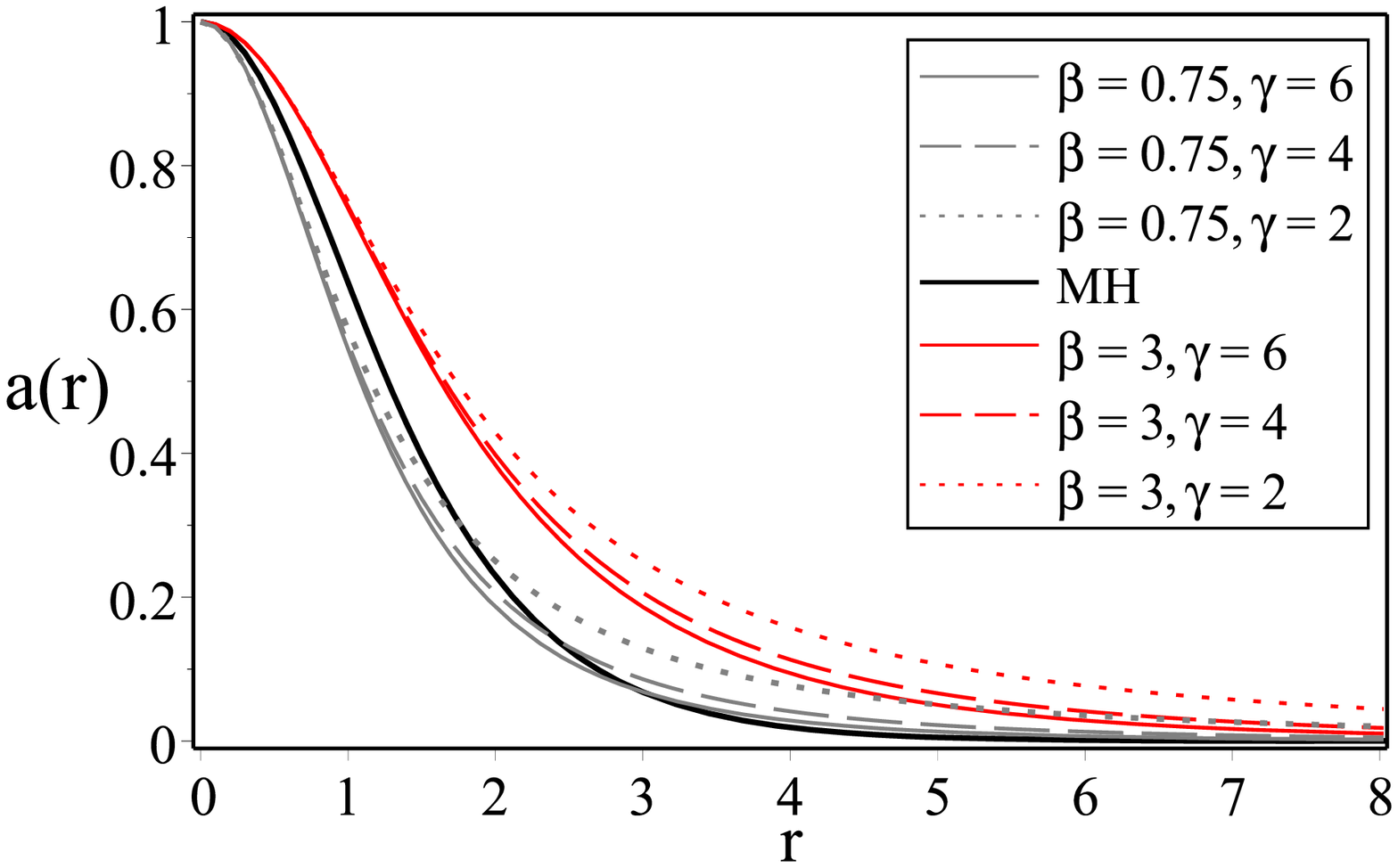}
\caption{The profiles $g(r) $ (upper) and $a(r) $ (lower) coming from the model
(\protect\ref{ss0}) with self-dual potential (\protect\ref{Vsd}).}
\label{fig05}
\end{figure}

Then, the BPS  equations describing the self-dual vortex solutions are
\begin{equation}
\frac{dg}{dr}=\pm \frac{ag}{r},  \label{bpsx1}
\end{equation}%
\begin{equation}
B=-\frac{1}{er}\frac{da}{dr}=\pm \frac{\beta }{2\beta -1}ev^{2}\left(
1-g^{2}\right) ^{1+2/\gamma}.  \label{bpsx3}
\end{equation}%
It is clear that the BPS equations of the usual Maxwell-Higgs model can be obtained
when $\beta =1$ and $\gamma\rightarrow\infty$.

The positive-definite self-dual energy density is
\begin{equation}
\varepsilon _{_{BPS}}=\frac{2\beta}{2\beta-1} V_{sd}(g)+ 2v^{2}\left( \frac{ag}{r}\right) ^{2}.  \label{denxt2}
\end{equation}

\begin{figure}[H]
\centering\includegraphics[width=8.6cm]{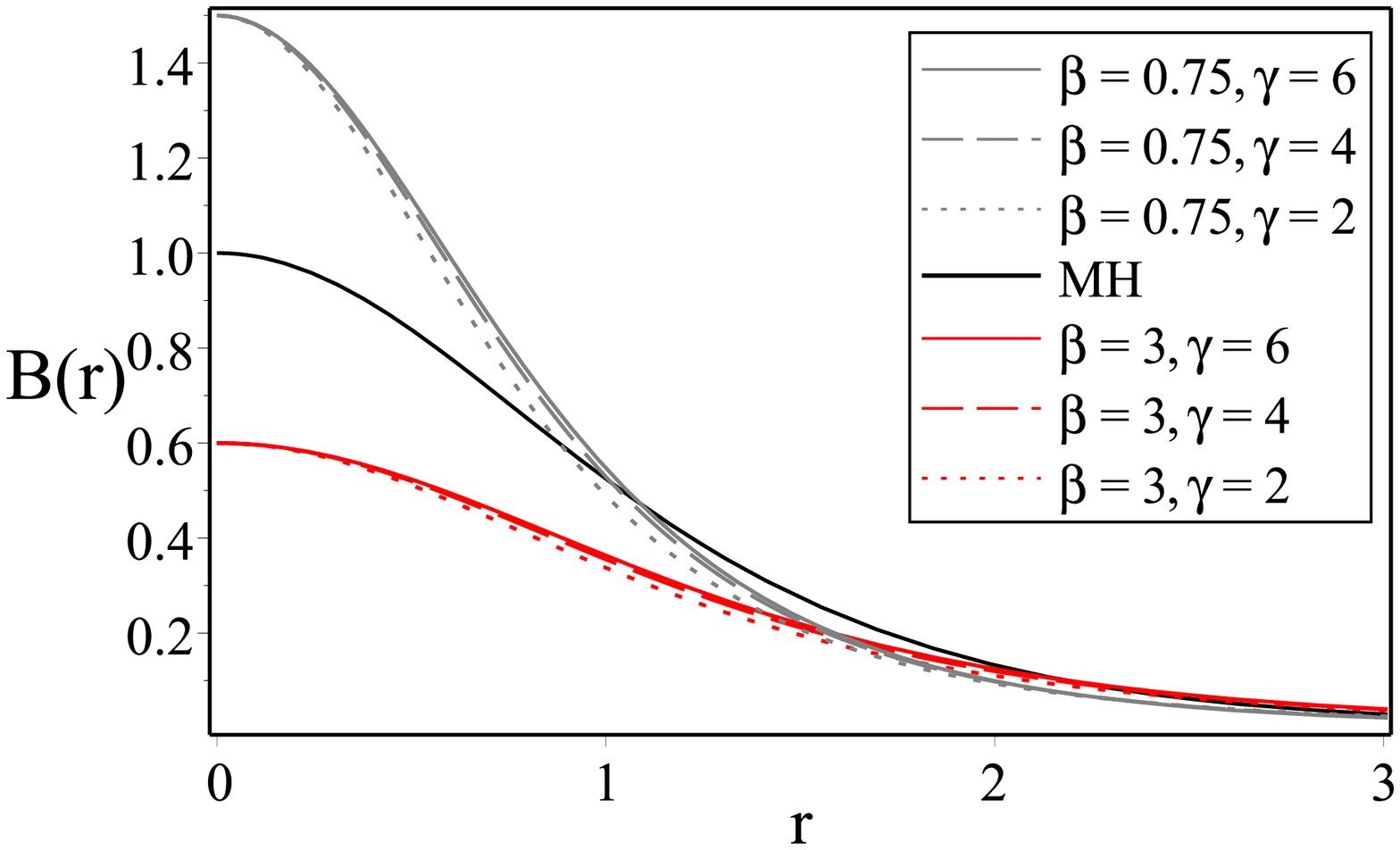}
\centering\includegraphics[width=8.6cm]{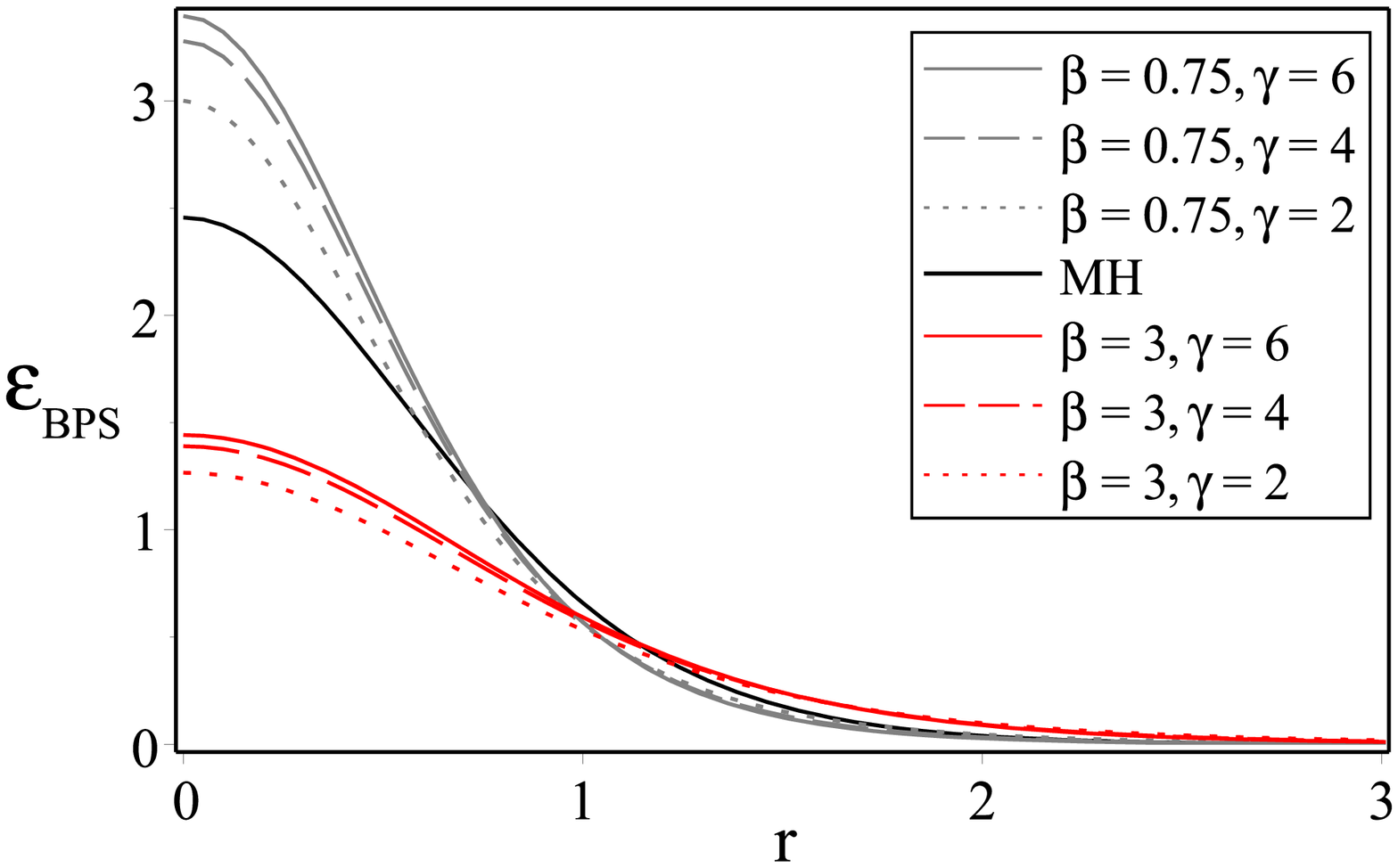}
\caption{The magnetic field $B(r) $ (upper) and the BPS energy density $%
\protect\varepsilon_{_{BPS}}(r) $ (lower) coming from the model (\protect\ref%
{ss0}) with self-dual potential (\protect\ref{Vsd}).}
\label{fig06}
\end{figure}

The behavior of $g(r)$ and $a(r)$ when $r\rightarrow 0$ is obtained
by solving the self-dual equations (\ref{bpsx1}) and (\ref{bpsx3}) around the
boundary values (\ref{bcx1}). Such analysis gives
\begin{eqnarray}
g(r) &\approx &G_{n}^{^{(\gamma )}}r^{\left\vert n\right\vert }+..., \\[0.15cm]
a(r) &\approx &n\mp  \frac{\beta }{2(2\beta -1)}\left( ev\right) ^{2}r^{2}+.....
\end{eqnarray}
We see that the behavior is similar to the localized vortices previously analyzed.

On the other hand, the solution of the BPS equations (\ref{bpsx1})
and (\ref{bpsx3}) for $r\rightarrow\infty$ provides a power-law decay for the
asymptotic behavior for the profiles  $g(r)$ and $a(r)$,
\begin{eqnarray}
g(r) &\simeq &1-\frac{1}{2}\left( \frac{\gamma ^{2}(2\beta-1)}{2\beta }\right) ^{\gamma
/2} \left( ev\right) ^{-\gamma}\frac{1}{r^{\gamma }}, \\[0.15cm]
a(r) &\simeq &\frac{\gamma }{2}\left( \frac{\gamma ^{2}(2\beta-1)}{2\beta }\right)
^{\gamma /2} \left( ev\right) ^{-\gamma }\frac{1}{r^{\gamma }}.
\end{eqnarray}%
It means that the vortex solutions are delocalized configurations
because their slow decay  for long distances in contrast with the
Abrikosov-Nielsen-Olesen vortices.

It is well know that the so-called London limit \cite{n013} provides
the behavior of the fields of a vortex in the full Ginzburg-Landau model which
is correctly predicts that the magnetic field varies monotonically and it is
exponentially localized at large distances. However, there are vortex solutions
having a  delocalized  magnetic field with profiles possessing slowly decaying.
These  delocalized vortices with  power-law decay has been obtained by studying
magnetic field delocalization in two-component superconductors \cite{Babaev}. Recently,
such a behavior has been also reported in diamagnetic vortices
generated within of a Chern-Simons theory \cite{xxzz}.

In order to compute the numerical solutions for the delocalized vortices,
we have select the upper signs, $e = v=1$ and the configurations with winding number
$n=1$. The numerical solutions of the BPS equations (\ref{bpsx1}) and (\ref{bpsx3})
have been performed for fixing $\beta=0.75$ (grey lines) and  $\beta=3$ (red lines).
For each $\beta$ we have selected some values of the power $\gamma=2,4,6$. The
resulting profiles for the Higgs and gauge fields are given in Fig. \ref{fig05}, the
correspondent ones for the  magnetic field and the self-dual energy density are
depicted in  Fig. \ref{fig06}.

\section{Ending comments\label{end}}

We have shown the existence of self-dual configurations in Abelian-Higgs
models where the kinetic term of gauge field is a highly nonlinear function of
$F_{\mu\nu}F^{\mu\nu}$. Our study is based in the Lagrangian density (\ref{lg1})
which in stationary regimen provides electrically neutral configurations. Starting from
the canonical energy density, we have implemented consistently the BPS formalism by
obtaining the general form of the self-interaction (potential) allowing to establish
that the total energy has a lower bound proportional to the magnetic flux.
Consequently, the field configurations having the minimum energy satisfy highly
nonlinear first-order differential equations the so called self-dual or BPS equations.
We have verified that the Lagrangian density (\ref{lg1}) besides to support self-dual
models in absence of a self-interacting potential also supports infinite twinlike
models with  $|\phi|^4$-potential or $|\phi |^6$-potential.

Among the infinite set of possible configurations, we have found families
of $k$-generalized models whose self-dual equations have a form mathematically similar
to the ones arising in the Maxwell-Higgs or Chern-Simons-Higgs models, i.e.,
$|D_{\pm}\phi|=0$ and $B=\pm \mathcal{V}(V_{sd}(|\phi|),h(|\phi|),{\cal{U}}(|\phi|)$,
where $\mathcal{V}$ is some function of the self-dual potential and generalizing
functions. The models with such a set of BPS equation could fulfill the condition
(\ref{cdx1}) despite the self-dual potential  (\ref{pot_x1}) present a complicated
form  but by choosing suitably the generalizing functions it is able to describe
$|\phi|^{4}$ or $|\phi|^{6}$ models. An example is the model defined by Eq. (\ref{ss0})
with potential given by Eq. (\ref{VSD}). Furthermore, in absence of a explicit
SSB potential, the condition (\ref{ccddt}) allows to describe simplest $k$-generalized
models admitting only $|\phi|^{4}$-self-interactions (see Eq. (\ref{sd_pt})).

With the aim to show explicitly that the BPS equations are able to
provide well-behaved solutions, we have considered the model given by Eq. (\ref{ss0})
to  study axially symmetric vortices. For a self-dual potential type $|\phi|^{4}$ or
$|\phi |^{6}$, we have shown that the $k$-generalized model is able to produce
solutions that for long distances have a exponential decay (as Abrikosov-Nielsen-Olesen
vortices). We have also shown that for the self-dual potential given by Eq.
(\ref{Vsd}), the vortices for long distances have a power-law decay (characterizing
delocalized vortices). They are remarkable solutions because such a
behavior has been obtained by studying magnetic field delocalization in two-component
superconductors \cite{Babaev} and recently in diamagnetic vortices \cite{xxzz}.
In all cases, we have observed that the generalization modifies the vortex-core size,
the magnetic field amplitude, the BPS energy density, the self-dual masses but the
total energy remains proportional to the quantized magnetic flux.

Finally, we consider two interesting challenges: Firstly to study the
influence of the $k $-generalized dynamics of the gauge field in the
existence of charged self-dual configurations in Abelian Higgs models with
Chern-Simons term or with Lorentz symmetry breaking terms. A second study is
to analyze cosmic string solutions for gauge field with $k $-generalized
dynamics in the context of modify gravity. Advances in this direction will
be reported elsewhere.

\begin{acknowledgments}
We thank CAPES, CNPq and FAPEMA (Brazilian agencies) for partial financial
support.
\end{acknowledgments}

\end{document}